\documentclass{article}[12pt]
\usepackage{amsmath,amssymb}
\usepackage{graphicx}%
\usepackage{verbatim}

\begin{document}

\begin{center}
{\bf \Large Energy-based approach to euclidean cycles in cosmological models
based on scalar fields\\[12pt]
Yu. G. Ignat'ev}\\
Physics Institute of Kazan Federal University,\\
Kremleovskaya str., 35, Kazan, 420008, Russia.
\end{center}

\begin{abstract}

On the basis of obtained equations of the energy balance for scalar
fields in cosmological models, a hypothesis formulated by the author on the
existence of Euclidean limit cycles in cosmological models based on scalar fields
with a Higgs type potential has been confirmed.  \\

{\bf keywords}: cosmological model, asymmetric scalar doublet, Euclidean
limit cycles.

\end{abstract}

\section{Basic equations of the cosmological model}

In [1--8] Ignat'ev and coauthors investigated cosmological models based on
classical and phantom scalar fields with potentials of Higgs type. In these
works, on the basis of qualitative and numerical analysis, the hypothesis was
expressed, of the possibility of the existence of so-called Euclidean limit
cycles to which cosmological evolution can tend.  In the present paper, on the
basis of obtained equations of energy balance, we bring clarity to this question.

In dimensionless variables, the closed system of ordinary differential equations
describing the cosmological evolution of an asymmetric scalar doublet in the case
of a spatially flat Universe has the form [1]
\begin{equation}
3\frac{{a{'^2}}}{{{a^2}}} = \left( {\Phi {'^2} + e{\Phi ^2} - \frac{{{\alpha
_m}}}{2}{\Phi ^4}} \right) - \left( {\varphi {'^2} - \varepsilon {\mu ^2}{\varphi
^2} + \frac{{{\beta _m}}}{2}{\varphi ^4}} \right) + {\lambda _m}
\label{eq:1}
\end{equation}
\begin{equation}
\Phi '' + 3\frac{{a'}}{a}\Phi ' + e\Phi  - {\alpha _m}{\Phi ^3} = 0
\label{eq:2}
\end{equation}
\begin{equation}
\varphi '' + 3\frac{{a'}}{a}\varphi ' - \varepsilon {\mu ^2}\varphi  + {\beta
_m}{\varphi ^3} = 0
\label{eq:3}
\end{equation}
where $e\;{\rm{and}}\;\varepsilon  =  \pm 1$,
\[
{\lambda _m} \equiv \frac{\lambda }{{{m^2}}},\,\,\,\,\,{\alpha _m} \equiv
\frac{\alpha }{{{m^2}}},\,\,\,\,\,{\beta _m} \equiv \frac{\beta
}{{{m^2}}},\,\,\,\,\,\mu  \equiv \frac{{\rm{m}}}{m},
\]
are $\alpha$ and $\beta$ tare the self-action constants for the classical and the scalar field, respectively,  
are the masses of these fields,  is the cosmological constant, and the prime denotes derivatives with respect to the dimensionless time variable $\tau$, connected with physical time  by the relation:
\[\tau=mt.\]
The large parentheses  in Eq. (1) are used to gather the contributions to the summed energy density from the classical field $\mathcal{E}_c$  and the phantom field $\mathcal{ E}_p$, respectively:
\begin{eqnarray}\label{einst}
\mathcal{E}_c=m^2\biggl(\Phi'\ \!^2+e\Phi^2-\frac{\alpha_m}{2}\Phi^4\biggr);&\displaystyle
\mathcal{E}_p=\biggl(-\varphi'\ \!^2+\varepsilon\mu^2-\frac{\beta_m}{2}\varphi^4\biggr),\nonumber\\
p_c=m^2\biggl(\Phi'\ \!^2-e\Phi^2+\frac{\alpha_m}{2}\Phi^4\biggr);&\displaystyle
p_p=\biggl(-\varphi'\ \!^2-\varepsilon\mu^2+\frac{\beta_m}{2}\varphi^4\biggr)\nonumber
\end{eqnarray}
\hspace{15pt}.

Introducing the potential energy of the classical field  and two phantom field
in the manner prescribed in [1]:
\begin{equation}\label{4}
V(\Phi)=
-\frac{\alpha}{4}\biggl(\Phi^2-e\frac{m^2}{\alpha}\biggr)^2; \quad
v(\varphi)=
-\frac{\beta}{4}\biggl(\varphi^2-\varepsilon\frac{\mathfrak{m}^2}{\beta}\biggr)^2,
\end{equation}
we write an expression for the normalized effective energy density:
\begin{equation}\label{5}
\mathcal{E}_m(\Phi,Z,\varphi,z)=\mathcal{E}_c+\mathcal{E}_p+\Lambda_m=\biggl(\frac{Z^2}{2}-V(\Phi)\biggr)-
\biggl(\frac{z^2}{2}-v(\varphi)\biggr)+\Lambda_m,
\end{equation}
where
\begin{equation}\label{6}
\Lambda_m=\lambda_m-\frac{1}{2\alpha_m}-\frac{\mu^2}{\beta_m}
\end{equation}
and introduce \textit{total normalized energies} of two classical and phantom
fields:
\begin{eqnarray}\label{7}
\mathcal{E}_c=\frac{Z^2}{2}-V(\Phi)=\frac{Z^2}{2}+\frac{\alpha}{4}\biggl(\Phi^2-\frac{e}{\alpha}\biggr)^2,\nonumber\\
\mathcal{E}_p=-\frac{z^2}{2}+v(\varphi)=-\frac{z^2}{2}-\frac{\beta}{4}\biggl(\varphi^2-e\frac{\varepsilon \mu^2}{\beta}\biggr)^2,\\
\mathcal{E}_m=\mathcal{E}_c+\mathcal{E}_p,\nonumber
\end{eqnarray}
connected with the standard values of the energies of these fields $E_c$  and $E_p$ by the
relations
\begin{equation}\label{8}
E_c=\frac{\mathcal{E}_c}{8\pi},\quad E_p=\frac{\mathcal{E}_p}{8\pi}.
\end{equation}
In this notation, Einstein's equation (Eq. (1)) can be rewritten in dimensionless form:
\begin{equation}\label{10}
\frac{a'\ \!^2}{a^2}\equiv H^2_m=\frac{1}{3}\mathcal{E}_m.
\end{equation}
In dimensionless variables, the closed normal bystem of ordinary differential
equations describing the cosmological evolution of an asymmetric scalar doublet
in the case of a spatially flat Universe has tme form [1]
\begin{eqnarray}
\Phi'=Z'\\
Z'=-\sqrt{6}Z\sqrt{\mathcal{E}_m(\Phi,Z,\varphi,z)}-e\Phi+\alpha_m\Phi^3,\\
\varphi'=z,\\
z'=-\sqrt{6}z\sqrt{\mathcal{E}_m(\Phi,Z,\varphi,z)}+\varepsilon\varphi-\beta_m\varphi^3.
\end{eqnarray}
\section{Equations of energy balance}
Multiplying both sides of Eq. (11) by $Z=\Phi'$ and both sides of Eq. (12) by $z=\varphi'$ with the
definition of the normalized energies of the fields (Eqs. (7)) taken into
account, we obtain the equations of balance for each of the fields:
\begin{eqnarray}\label{14}
\frac{d\mathcal{E}_c}{d\tau}=-\sqrt{6}Z^2\sqrt{\mathcal{E}_m},\\
\frac{d\mathcal{E}_pc}{d\tau}=\sqrt{6}z^2\sqrt{\mathcal{E}_m}.
\end{eqnarray}
It follows from these equations that for $\mathcal{E}_m(\Phi,Z,\varphi,z)>0$ the total energy of the classical
scalar field can only decrease with time whereas the total energy of the phantom
field can only increase.  Adding the two sides of Eqs. (14) and (15), the left
side to the left side and the right side to the right side, we obtain the
equation of energy balance:
\begin{eqnarray}\label{16}
\frac{d\mathcal{E}_m}{d\tau}=-\sqrt{6}(Z^2-z^2)\sqrt{\mathcal{E}_m}\Rightarrow \frac{d\sqrt{\mathcal{E}_m}}{d\tau}=-\sqrt{\frac{3}{2}}(Z^2-z^2)\nonumber\\
\Rightarrow \sqrt{\mathcal{E}_m(\tau_0)}-\sqrt{\frac{3}{2}}\int\limits_{\tau_0}^\tau (Z^2-z^2)d\tau.
\end{eqnarray}
A very interesting fact follows from Eq. (16), namely, that for $|Z|>|z|$\footnote{That
is to say, in the case in which the classical scalar field dominates.} the
effective energy falls with time, tending to the limit $\mathcal{E}_m\to0$. In the opposite case in
which the phantom field dominates ($|Z|<|z|$) the effective energy can only grow with time,
which appears to lead  to an instability of the cosmic model.

\section*{Concluisons}

The obtained result is found to be in agreement with the results of qualitative
and numerical modeling of cosmological models, obtained by the author and his
coauthors, for example, in [2--8]. In particular, in the case of an isolated
classical scalar field, when Em. (14) can be rewritten in the form
\begin{equation}
\varphi\equiv0 \Rightarrow \mathcal{E}_m=\mathcal{E}_c+\Lambda_m\Rightarrow \frac{d\mathcal{E}_m}{d\tau}=-\sqrt{6}Z^2\sqrt{\mathcal{E}_m}.
\end{equation}
It follows from Eq. (17) that in the case of an isolated classical field with a
Higgs interaction potential, over the course of time, the effective energy tends
to zero, i.e., the Universe tends to a Euclidean Universe.   Here, two cases are
possible: 1) as $\mathcal{E}_m\to 0$, $\Phi^2+Z^2\to 0$  and $\Lambda_m=0$; and 2) as $\mathcal{E}_m\to 0$, $\Phi^2+Z^2\not= 0$  and $\Lambda_m$  can take any value.  As was shown in the cited works, the second case can be realized, and in this case the scalar
field remains nonzero, completing oscillations supporting dynamical equilibrium.

\end{document}